\documentstyle[preprint,eqsecnum,aps,prbbib]{revtex}
\tightenlines

\def\cm{~cm$^{-1}$}
\def\hx{HX}
  
\def\hos{$H1s$}
\def\hts{$H2s$}  
\def\htpp{$H2p^+$}
\def\htpm{$H2p^-$}
 
\def\los{$L1s$} 
\def\ostpp{\hos$\rightarrow$\htpp}

\def\tpmts{\htpm$\rightarrow$\hts}

\def\com{_{\mathrm{\tiny COM}}}

\begin{document}
\title{Photo-thermal transitions of magneto-excitons in quantum wells}

\author{J. \v Cerne,$^{1*}$ J. Kono,$^{2*}$ M. Su,$^3$ and M.S. Sherwin$^4$}

\address{$^1$ Department of Physics, University at Buffalo, The State University of New York , Buffalo, NY 14260, USA.
}
\address{$^2$ Department of Electrical and Computer
 Engineering, Rice University, Houston, TX 77005, USA.
}
\address{$^3$ National Institutes of Standards and Technology 815.04
325 Broadway, Boulder, CO  80305, USA.
}
\address{$^4$  Department of Physics and Center for Terahertz Science and Technology,\\ 
University of California, Santa Barbara, California 93106, USA.
}

\date{to be submitted to Phys. Rev. B 6/13/02}
\maketitle

\begin{abstract} 
By monitoring changes in excitonic photoluminescence (PL)  that are
induced by terahertz (THz) radiation, we observe resonant THz absorption
by magnetoexcitons in GaAs/AlGaAs quantum wells.  Changes in the PL spectrum are explored 
as a function of temperature and magnetic field, providing insight into the mechanisms 
which allow THz absorption to modulate PL.  The strongest PL-quenching occurs at the 
heavy hole $1s\rightarrow 2p^+$ resonance where heavy hole excitons are photo-thermally 
converted into light hole excitons. 
\end{abstract} 
\pacs{78.66.-W, 78.30.Fs, 71.35.Ji}

\narrowtext
 
Undoped quantum wells (QW) have been studied extensively using interband
optical techniques (for examples, see
Ref.~\onlinecite{seiler}).  These experiments have revealed a rich
structure which is dominated by correlated electron-hole pairs known as
excitons.  Excitons in square QW form hydrogen-like states with binding
energy of approximately 10 meV (the energy of a terahertz photon) and Bohr radius of 100 \AA.  
The internal dynamics of excitons can be explored by using visible or near-infrared photons to
create excitons, and THz radiation (of order 80 \cm, 10 meV, 2.4 THz) to
manipulate them directly.  At low THz intensities, one expects direct
transitions between even- and odd-parity states of the exciton which are not observable 
with linear interband spectroscopy.  Such transitions provide new, sensitive tests for 
the theory of excitons, which is fundamental in the physics of semiconductors.  At higher 
THz intensities, one also can probe the dephasing and energy relaxation rates of excitons. 

Undoped direct gap (Type I) QW are especially interesting since they are
so commonly used and provide a simple model system for theoretical
analysis.  However, the short lifetime of excitons in Type I QW makes it
difficult to achieve the large population of cold excitons required for
THz absorption studies.  Experimental progress has been made in QW using
photoinduced absorption in staggered (Type II)  QW\cite{hodge} and
time-resolved THz spectroscopy in Type I QW.\cite{grischkowsky} Optically-detected THz 
resonance (ODTR) spectroscopy was used
to probe direct excitations of internal transitions of
magnetoexcitons in Type I QW for the first time.\cite{cerne2,salib} More recent work has included the 
study of charged\cite{nickel,dzyubenko2,kios} as well as neutral\cite{nickel2,herold} 
magnetoexcitons in Type I QW.

Though ODTR has proven to be a powerful and successful technique, it
remains an indirect measurement of THz absorption which still is not 
well understood.  In this paper we use a multichannel technique to capture the full 
PL spectrum as a magnetic field sweeps exciton energy levels through THz
resonances.  We find a combination of resonant and quasi-thermal excitations are 
responsible for THz-induced changes in PL.  In a particularly striking resonant 
process, THz radiation converts  heavy-hole excitons (HX) to (radiative) light-hole excitons (LX).  

The undoped sample consists of fifty 100 \AA-wide GaAs square QW between 150
\AA-wide Al$_{0.3}$Ga$_{0.7}$As barriers.\cite{quinlan}  UCSB's Free-Electron Lasers 
(FEL) provide intense radiation (up to
1~MW/cm$^2$) that can be continuously tuned from 4 to 160 \cm\ 
(0.5-20 meV, 0.12-4.8 THz).  As shown in Fig.~\ref{fig;exset}, visible radiation from an Ar$^+$ laser
is used to create electron-hole pairs in the undoped sample. 
Simultaneously, THz radiation with the electric field polarized in the
plane of the QW, and therefore not coupling to intersubband transitions,
passes through the sample.  Measurements are performed with a visible laser excitation 
intensity of approximately 100~Wcm$^{-2}$,
creating an exciton density of approximately $3\times10^{10}$~cm$^{-2}$
per well in the sample.\cite{density} The resulting PL is captured by
eighteen 50~$\mu$m diameter optic-fibers that surround a central
fiber used for excitation.  The PL is delivered to a monochromator and
detected by an image-intensified charge-coupled device (ICCD) camera.  The output of the Ar$^+$
laser is modulated acousto-optically to produce a 20 $\mu$s visible
excitation pulse that coincides with the 5 $\mu$s THz pulse at the sample. 
The ICCD is gated in time to capture the entire PL spectrum while the THz
radiation pulse illuminates the sample. Since the THz pulse is much longer than any
carrier relaxation time, the measurement is in steady-state.  The PL
change during the THz pulse is due to carrier excitation; no lattice heating
is observed.\cite{cerne1} 

The measurement involves monitoring the PL spectrum as a function of magnetic field $B$
while fixed frequency THz radiation illuminates the sample.  Unlike a single channel 
measurements using a photo-multiplier tube, the ICCD allows the entire PL spectrum to 
be captured at each value of $B$.  This powerful new technique is referred to as multichannel 
optically-detected THz resonance (MOTR). A separate MOTR run is made with the THz 
radiation blocked to serve as a
reference.

Figure~\ref{fig;mplspectra} shows the difference of the PL
spectra with and without THz irradiation plotted as a
function of $B$. The THz frequency is 103\cm\ and the intensity is approximately 50 kW/cm$^2$.  The sample 
temperature is 10~K. Stronger PL quenching is represented by more negative values 
(shown in blue) of the PL difference.   The false color image of this surface plot is 
projected on the bottom of the graph.  There are strong qualitative differences in the PL
modulation at different magnetic fields.  Since experiments begin from $B=0$~T, it is convenient to label the excitonic states using a hydrogenic notation (principal quantum number, orbital angular momentum, z-projection of the orbital angular momentum).  At $B=0$~T, the difference spectrum
shows the quenching of the heavy hole $1s$ (\hos) PL peak amplitude (labeled A in
Fig.~\ref{fig;mplspectra}), the enhancement of high energy tail of the
\hos\ PL (B), and the enhancement of the light hole $1s$ (\los) PL amplitude (C).  These
effects were previously observed and studied in detail.\cite{cerne1} As
$B$ increases, the \hos\ PL amplitude is resonantly quenched at the
\ostpp($m_z=+1$) transition at 3.5~T (D and D'), 
while the \los\ PL reaches a
maximum (E and E').  The
THz-induced transitions from \hos\ to  \los\ states is clearly seen. 
At 8 T, further resonant quenching is observed (F) which is assigned to the $H2p^-(m_z=-1)\rightarrow H2s$ transition.
Note that above
3.5~T, quenching is reduced and the dominant effect of THz
radiation is to enhance and blueshift the \hos\ PL peak (F). This can seen in the depth 
(height) of the PL difference surface plot above and below 3.5~T and is is also shown by the 
color difference between these two regions in the projected image plot. A ridge that 
begins at 1.58~eV at 0~T and appears to follow the quadratic magnetic field dependence 
of the \hos\ PL peak also is indicated (G).  This ridge is consistently higher than 
the \hos\ state by one THz photon energy throughout the magnetic field range.  Note 
that this ridge appears to cross a more strongly curved second ridge at point E', 
which is where the \los\ is resonantly enhanced. Neither ridge is consistent with 
magnetic field dependence of PL from the \los.  Also note that the line where the 
PL difference sheet crosses zero (yellow line in the projected image plot) is nearly 
independent of magnetic field and does not follow the $B^2$ dependence of 
\hos\ PL peak energy.

The information contained in Fig.~\ref{fig;mplspectra} is more accessible
when cross-sections through this surface plot are examined.  The resonant
enhancement of the \los\ PL peak at the expense of the \hos\ PL amplitude
is shown in Fig.~\ref{fig;ampsvsB}(b), where the \hos\ and \los\ PL
amplitudes are plotted against $B$.  Note that this figure shows the PL amplitudes, 
not the normalized PL ratios (i.e., PL with divided by PL without THz irradiation, as shown
in most ODTR spectra), which adds a
baseline that depends on $B$.  The $H1s\rightarrow H3p$ feature at 2.1~T 
splits into two dips ($H1s\rightarrow H3p^+$ and $H1s\rightarrow H3p^-$) 
above 2.5~T.\cite{cerne2,salib,nickel00} 
The \los\ PL
enhancement is asymmetric in $B$, with a fast decay above 3.5~T, and a more gradual decrease
below 3.5~T.  The \hos\ amplitude shows similar asymmetry and reflects the 
conventional ODTR results.\cite{cerne2,salib} Figure~\ref{fig;ampsvsB}(a)
sketches the energy difference between the \hos\ and higher energy exciton states (based on data 
published in Ref.~\onlinecite{cerne2}).  The diagrams in the inset of Fig.~\ref{fig;ampsvsB}(a) sketch the photo-thermal carrier distributions at different magnetic fields, and will be discussed in greater detail later in this paper.

Absorption of THz radiation can heat carriers, resulting in a quasi-thermal distribution with a 
temperature above that of the lattice.\cite{cerne1}  It is thus useful to compare changes in 
photoluminescence induced by THz radiation to those induced by changing the temperature of the sample.  
Figure~\ref{fig;arrhen} plots the \los\ to \hos\ 
PL amplitude ratio as a function of inverse lattice temperature, without (a) and with (b) THz 
irradiation of approximately 100~Wcm$^{-2}$.  These data form Arrhenius plots where straight lines indicate thermally activated 
population of the \los\ from \hos\ states.  The slopes of these lines determine the activation 
energies, which are shown in the legends of Fig.~\ref{fig;arrhen}.  Note the dramatic decrease in the activation energy when the sample is illuminated by THz radiation at 3.5~T.

We propose a qualitative explanation of THz-induced changes in PL which involves 
three processes.  
1. Absorption:  THz radiation is absorbed via internal 
transitions of  excitons,\cite{cerne2,salib,nickel00,nickel,herold} 
or by Drude-like free carrier excitations of ionized electrons and holes.\cite{cerne1}  
2. Quasi-thermalization:  Each species of excitons (for example, \hos, \htpp, or \los), 
is associated with a band of states with different center-of-mass momenta.  The power transferred 
into the electron-hole system indirectly heats each band of excitons to a temperature warmer than 
that of the lattice.   
3. Luminescence:  The luminescence under THz irradiation only reflects the population of states that can participate in 
radiative interband transitions.  In bands with even parity ($1s$, $2s$, for example), the radiative states 
are those with in-plane momentum $K\com$ near zero.  In bands with odd parity, such as $2p$, no states 
are radiative.

We now discuss the experimental data as a function of increasing magnetic field with a fixed THz frequency of 103~\cm.  The case $B=0$~T 
has been studied extensively in Ref.~\onlinecite{cerne1}.  In this nonresonant case, the mechanism for absorption 
was shown to be Drude-like heating of ionized electrons and holes.  These hot electrons and holes 
heat luminescing excitons.  The non-resonant photo-thermal distribution is represented by the cartoon on the left side of the inset in Fig.~\ref{fig;arrhen}(a). Under THz illumination, the exciton system (including both \hos\ and 
\los\ excitons) could be reasonably well described by a single  temperature larger than that of the 
lattice.  This temperature could be measured by matching the PL spectrum with THz irradiation at a 
low lattice temperature to the PL spectrum without THz radiation at an elevated lattice temperature.  
As either temperature or THz power increases, the magnitude of \hos\ luminescence decreases, and the 
\los/\hos\ PL ratio increases.  Either of these quantities could be used to measure the temperature of 
the exciton system.  Both measures produce consistent results.\cite{cerne1}  

Between 0~T and about 3.5~T, the qualitative features of Figs. 2 and 3 are similar to those at 
B=0, with a quenching of the \hos\ luminescence and enhancement of \los.  In this regime, there 
are many internal states which have energies below that of the 103~\cm\  pump, and which can 
be excited by it.  

One of the most striking features of the data is the resonant transfer of PL from \hos\  to \los\  
which occurs near 3.45~T (E and E' in Fig.~\ref{fig;mplspectra}, also 
Fig.~\ref{fig;ampsvsB}).  Our explanation is a resonant 
photo-thermal mechanism, shown schematically in the central cartoon in the inset of Fig.~\ref{fig;arrhen}(a).  THz radiation promotes excitons 
from \hos\ to \htpp.  The dark \htpp\ state\cite{single} is only about 1~meV below 
the radiative \los\  state according to PL and ODTR data.  Thermal fluctuations, associated with the quasi-temperature of 
the \htpp\ excitons, are sufficient to then populate \los\ exciton states to a much 
greater degree than they could be populated via thermal excitation directly from \hos.  According to photo-thermal ionization experiments on shallow donors in
GaAs,\cite{stillman} the $2p^+$ state allows extremely efficient
ionization.  Reference~\onlinecite{stillman} found that the probability of
ionization from the $2p$ state is much larger than expected from the
energy separating this state and the continuum; the ionization probability at 4.2~K
is essentially unity.\cite{above} \los\ 
excitons decay, emitting the observed luminescence. The photo-thermal hypothesis is 
supported by the observation that, in Fig.~\ref{fig;arrhen}(b), the ratio of \los\  to \hos\  luminescence is 
nearly independent of temperature for the case of THz on, B=3.45~T.  This demonstrates 
an effective activation energy which is negligible.  In contrast, without THz irradiation, the \los/\hos\ at 3.45~T increases strongly with temperature, consistent with an 
activation energy on the order of 10~meV.

A second hypothesis is that the THz radiation heats the entire exciton system efficiently on 
resonance, but that all bands of excitons can still be described with a single temperature.  
We call this the thermal hypothesis.  The thermal hypothesis could be consistent with the 
insensitivity of the \los\ /\hos\  ratio to increasing temperature at resonance.  However, it predicts that, 
as was observed for B=0~T, the photoluminescence spectrum under THz illumination at low 
temperatures can be mapped onto the photoluminescence spectrum without THz excitation at 
some higher temperature.  This is not the case.  At 3.45~T and at a lattice temperature of 10~K with THz 
illumination of 110~kW/cm$^2$, the \hos\  PL amplitude is the same as for a lattice temperature of 70~K without 
THz illumination at 3.45~T.    However, the \los\  peak for the cold sample illuminated with THz is over 40\% larger than for the unilluminated warm sample.  Furthmermore, the \los\ PL amplitude increases linearly with THz intensity at the \ostpp\ resonance, while the increase in the absence of THz illumination is consistent with exponentially activated behavior as a function of temperature.  This is consistent with the photo-thermal hypothesis, and inconsistent with the thermal hypothesis.

Above 3.45 T, the changes in photoluminescence induced by THz irradiation are qualitatively 
different from those induced at lower magnetic fields.  At high magnetic fields in
Fig.~\ref{fig;ampsvsB}(a), the THz photon energy is smaller than the
lowest dominant exciton transition (\ostpp).  The high magnetic field photo-thermal distribution is shown schematically in the cartoon on the right side of the inset in Fig.~\ref{fig;ampsvsB}(a).  In this regime, lower energy
exciton transitions (e.g., \tpmts) can occur, and the hot carriers can
impact with and heat cold excitons, which produces a PL quenching signal
at 8~T in Fig.~\ref{fig;mplspectra}.  The heating is not efficient and
high THz and visible intensities are required to produce the \tpmts\
feature.    There are several
reasons for this reduced efficiency.  First, at low temperature, most of
the carriers are bound in $1s$ (ground state)  excitons, so the population
in excited states (e.g., $2p^-$ which has an energy that has a characteristic temperature 
of 80~K above the $1s$ state), especially at low THz and visible
intensities, is significantly lower than the $1s$ population.  Higher
intensities increase the exciton temperature, and hence enhance the
population of $2p^-$ excitons that can participate in the \tpmts\
transition.  Second, PL amplitudes are not as sensitive to temperature changes at higher
$B$.   Finally, most 
hot heavy hole excitons (\hx) (with center of mass momentum
$K\com\ne0$) will eventually cool to contribute to the main \hx\ PL line within 
the radiative cone ($K\com\approx0$).  Since radiative recombination 
for the hot \hx\ ($K\com\ne0$) is
suppressed due to momentum conservation, and since non-radiative
recombination is weak, most of the hotter \hx\ have nowhere to go but
back to the $K\com=0$ state where they can efficiently recombine
radiatively.  The closest escape for \hx\ is the \los\ which is over 14
meV (163~K) higher in energy at $B=0$ (see Fig.~\ref{fig;ampsvsB}(a)). This
activated behavior can be seen in Fig.~\ref{fig;arrhen}, where the
activation energy to populate the \los\ state is approximately 10 meV
without (a)  and with (b) THz irradiation at 8~T.  As a result, heating
may increase the \hos\ lifetime as it journeys through $K\com$-space, but will
not significantly change the quantum efficiency of the \hos\ PL nor the
steady-state ODTR signal. 
 
This model does not explain the THz-induced enhancement and blueshift of
the \hos\ PL amplitude.  One expects an energy
shift in the PL close to THz resonances due to the ac Stark effect.\cite{stark}
The calculated shift changes sign depending on whether the THz
frequency is higher or lower than the transition frequency.  The data, 
on the other hand, show a shift that increases monotonically with
$B$. The lack of a sign change may be due to the fact excitonic transition energies 
are swept across the energy of the photons in the strong THz radiation field.  
As a result the detuning frequency is not fixed, but is swept from positive to 
negative values, and hence no single laser induced energy shift is observed.

MOTR has been used to study the THz dynamics of magnetoexcitons
in GaAs QW.  Excitonic transitions have been identified and are in good
agreement with theory.  A simple qualitative model can be used to
understand the PL modulation due to THz radiation.  On the other hand,
this work has generated several new puzzles which still require exploration.  The MOTR baseline
and the higher energy ridge (labeled G in Fig.~\ref{fig;mplspectra}) that follows the \hos\ peak are still not clearly understood. The higher energy ridges (G and E' in Fig.~\ref{fig;mplspectra}) appear to indicate mixing of excitonic and photonic states. Significant energy shifts of over 3 meV are observed in the
\hos\ PL close to \tpmts, which may be related to an ac Stark effect.  The
information gained from studying excitons in QW will provide an important
foundation for research of excitons in even lower dimensional systems such
as quantum wires and dots. 

\acknowledgments The authors gratefully acknowledge the assistance of T.
Gutierrez, G.E.W. Bauer and A.B. Dzyubenko.  Samples were kindly provided by M. Sundaram and A. C. Gossard.  We would also
like to thank D.P. Enyeart and J.R. Allen at the Center for Terahertz Science and Technology for their technical support.  This work has been supported
by the NSF Science and Technology Center for Quantized Electronic
Structures DMR 91-20007, NSF-DMR 0070083, ONR N00014-K-0692, AFOSR 88-0099, and the Alfred P. Sloan Foundation (MSS).

$^{*}$ All measurements were performed at the Center for Terahertz Science and Technology, University of California, Santa Barbara, California.

\begin{figure}
\caption{A schematic of the experimental setup for MOTR measurements.  
Photoluminescence is detected while the THz pulse passes
through the sample.  The THz radiation is polarized parallel to the plane 
of the QW.  The timing of the laser pulses and PL is shown in
the lower half of the figure.} 
\label{fig;exset}
\end{figure}

\begin{figure}
\caption{PL spectra without THz irradiation subtracted from
PL spectra with THz irradiation at 103\cm\ as function of $B$. 
The THz intensity is approximately 50 kWcm$^{-2}$.  The sample temperature is 10~K.}
\label{fig;mplspectra}
\end{figure}

\begin{figure}
\caption{Energies relative to the \hos\ state of higher energy excitonic states are 
schematically shown in (a).  Solid lines indicate energy levels (e.g., \htpp) that 
are accessible from the 
\hos\ state via radiative transitions.  When $B=B_{\hbox{1s-2p$^+$}}$, the THz photon energy 
matches the \ostpp\ energy separation.   Note that for
$B>3.5$~T, the \htpp\ state is above the \los\ state. Inset in (a) are simple 
representations of the thermal and photo-thermal carrier distributions induced by 
THz irradiation. The \los\ and \hos\ PL amplitudes are plotted as a function of $B$
under THz illumination at 10~K in (b).  These data are obtained from cross-sectional 
slices that follow the \los\ and \hos\ PL energies in Fig.~\ref{fig;mplspectra}. }
\label{fig;ampsvsB}
\end{figure}

\begin{figure}
\caption{Arrhenius plot of the \los/\hos\  ratio without (a) and with (b) 
THz irradiation.  The THz intensity is approximately 100~kWcm$^{-2}$. The straight 
lines represent activated
behavior with an activation energy that is determined by the slope of the line.  
Activation energies that are obtained from exponential fits are shown in the legends.  The
\los-\hos\  energy spacing measured from PL is approximately 14
meV for all $B$.  The \htpp-\los\ separation is
approximately 1 meV at 3.45~T.}
\label{fig;arrhen} 
\end{figure}


\begin{references}
\bibitem{seiler}{\it The Spectroscopy of 
Semiconductors}, Semiconductors and Semimetals {\bf 36},
eds. D.G. Seiler and D.L. Littler, (Academic Press, New York, 1992).

\bibitem{hodge}C.C. Hodge, C.C. Phillips, M.S. Skolnick, G.W. Smith, 
C.R. Whitehouse, P. Dawson, and C.T. Foxon, Phys. Rev.
B {\bf 41}, 12319 (1990).

\bibitem{grischkowsky}R.H.M. Groeneveld and D. Grischkowsky, J. Opt. 
Soc.  Am. {\bf 11}, 2502 (1994).

\bibitem{cerne2}J. \v Cerne, J. Kono, M.S. Sherwin, M. Sundaram, 
A.C. Gossard, and G.E.W. Bauer, Phys.
Rev. Lett. {\bf 77}, 1131 (1996).

\bibitem{salib}M.S. Salib, H.A. Nickel, G.S. Herold,A. Petrou, B.D. McCombe,
R. Chen, K.K. Bajaj, and W. Schaff,  Phys. Rev. Lett {\bf 77}, 1135 (1996). 

\bibitem{nickel}H.A. Nickel, G.S. Herold, T. Yeo, G. Kioseoglou, Z.X. Jiang, B.D. McCombe, A. Petrou, D. Broido, and W. Schaff, Phys. Status Solidi B {\bf 210}, 341 (1999).

\bibitem{dzyubenko2}A.B. Dzyubenko, A.Y. Sivachenko, H.A. Nickel, T.M. Yeo, G. Kioseoglou, B.D. McCombe, and A. Petrou, Physica E {\bf 6}, 156 (2000).

\bibitem{kios}G. Kioseoglou, H.D. Cheong, T. Yeo, H.A. Nickel, A. Petrou, B.D. McCombe, A.Y. Sivachenko, A.B. Dzyubenko, and W. Schaff, Phys. Rev. B {\bf 61}, 5556 (2000).

\bibitem{nickel2}H.A. Nickel, G. Kioseoglou, T. Yeo, H.D. Cheong, A. Petrou, B.D. McCombe, D. Broido, K.K. Bajaj, and  R.A. Lewis,
Phys. Rev. B {\bf 62}, 2773 (2000).

\bibitem{herold}G.S. Herold, H.A. Nickel, J.G. Tischler, B.A. Weinstein, and B.D. McCombe, Physica E {\bf 2}, 39 (1998).

\bibitem{quinlan}S.M. Quinlan, A. Nikroo, M.S. Sherwin, M. Sundaram, 
and A.C. Gossard, Phys. Rev. B {\bf45}, 9428 (1992).

\bibitem{density}The exciton density is determined using the
absorption coefficient for GaAs at 532 nm of $8\times10^4$~\cm\ and
an exciton lifetime of 0.5 ns.

\bibitem{cerne1}J. \v Cerne, A.G. Markelz, M.S. Sherwin, S.J. Allen,
M. Sundaram, A.C. Gossard, P.C. van Son, and D. Bimberg, Phys. Rev.
B {\bf 51}, 5253 (1995).

\bibitem{nickel00}H. A. Nickel, G. Kioseoglou, T. Yeo, H. D. Cheong, 
A. Petrou, B. D. McCombe, D. Broido, K. K. Bajaj, and R. A. Lewis, Phys. Rev. B {\bf 62}, 2773 (2000)

\bibitem{single}Single photon radiative recombination is forbidden for the $2p^+$ state, and non-radiative recombination is also weak.

\bibitem{stillman}G.E. Stillman, C.M. Wolfe, and D.M. Korn,
Proceedings of the International Conference on the Physics of
Semiconductors, Warsaw, Poland, 1972.

\bibitem{above}Above approximately 3 T, the  \htpp\  is in fact
higher in energy than the \hx\  continuum.

\bibitem{stark}For references discussing the THz Stark effect in
semiconductor quantum heterostructures see B. Birnir, B. Galdrikian, R.
Grauer and M. Sherwin, Phys. Rev. B {47}, 6795 (1993) and M.S. Sherwin
in ``Quantum Chaos,'' Edited by G. Casati and B.V. Chirikov, p. 209, 1995, p. 209, Cambridge University Press.

\end{references}
\end{document}